\documentclass[a4paper,12pt]{article}
\usepackage[a4paper,text={16.8cm,22.4cm}]{geometry}
\usepackage{amsmath,amssymb,latexsym,bm,braket}
\usepackage{cite}
\usepackage{graphicx}
\usepackage{color}

\allowdisplaybreaks 
\newcommand{\Tr}{\mathrm{Tr}}

\newcommand{\ff}{f\hspace{-0.4em}f}
\newcommand{\fff}{f\hspace{-0.2em}f}

\newcommand{\be}{\begin{equation}}
\newcommand{\ee}{\end{equation}}
\newcommand{\bea}{\begin{eqnarray}}
\newcommand{\eea}{\end{eqnarray}}

\begin{document}

\begin{titlepage}

  \begin{flushright}
  IPPP/13/35
  \end{flushright}
  
  \vspace{5ex}
  
  \begin{center}
    \textbf{ \Large Top-quark pair production at high invariant mass: an NNLO soft plus virtual approximation}

    \vspace{7ex}
    
    \textsc{Andrea Ferroglia$^b$, Ben D. Pecjak$^c$, and Li Lin Yang$^{a,d}$}

    \vspace{2ex}
  
    \textsl{
${}^a$School of Physics and State Key Laboratory of Nuclear Physics and Technology\\
Peking University, Beijing 100871, China
\\[0.3cm]
${}^b$New York City College of Technology, 300 Jay Street\\
Brooklyn, NY 11201, USA
\\[0.3cm]
${}^c$Institute for Particle Physics Phenomenology, University of Durham\\
DH1 3LE Durham, UK
 \\[0.3cm]
${}^d$Center for High Energy Physics, Peking University, Beijing 100871, China
    }
  \end{center}

  \vspace{4ex}

  \begin{abstract}

We obtain a soft plus virtual approximation to the NNLO QCD contributions to the top-pair invariant mass distribution at hadron colliders.  It is valid up to corrections of order $m_t^2/M^2$, with $M$ the pair invariant mass. This is currently the most complete QCD calculation for a differential cross section in top-quark pair production, and is useful for describing the high invariant mass region characteristic of boosted top quarks.  We use our results to construct an improved NNLO approximation for the pair invariant mass distribution and compare it with previous, less complete approximations based on logarithmic terms from NNLL soft-gluon resummation alone. We find that the new NNLO approximation produces moderate enhancements of the differential cross section compared to previous ones, the effect being slightly more important at low values of invariant mass than at high ones. On the other hand, at high values of invariant mass the new NNLO corrections are dominated by even higher-order effects included in NNLL soft-gluon resummation, reaffirming the need for resummation in describing the highly boosted regime.

\end{abstract}

\end{titlepage}

\section{Introduction}
\label{sec:intro}

The production of top-quark pairs is a process of primary importance in elementary particle physics. Top quarks are unique among quarks in the fact that they decay before hadronizing. Furthermore, because of their very large mass, top quarks are expected to play a key role in the study of the mechanism of electroweak symmetry breaking. Several observables related to top-quark physics have been measured with great accuracy at the Tevatron, where the top quark was first discovered in 1995. In particular, the top-quark mass was determined with a relative error of less than one percent, while the total top-quark pair production cross section was measured with a relative error of less than 10\%. Some pair production differential distributions were also measured at the Tevatron, albeit with larger errors. These include the top-quark transverse momentum distribution, the top-pair invariant mass distribution, and the top quark forward-backward asymmetry (whose measured value is significantly larger than the corresponding Standard Model prediction). The large number of top-quark events observed at the Large Hadron Collider (LHC) will allow for precise measurements of some of these differential distributions. In particular, the ATLAS collaboration already measured the top-quark pair invariant mass distribution up to about 2~TeV for a center-of-mass energy of 7~TeV \cite{Aad:2012hg}.

Several conjectured frameworks for physics beyond the Standard Model predict the existence of new particles which can decay into highly boosted top quarks. If one of these models is realized in Nature, one would expect to observe resonant bumps or more subtle distortions in the pair invariant mass distribution, especially at large values of the pair invariant mass. The pair invariant mass distribution can then be used to set bounds on the allowed parameter space of such models, including some of those proposed in order to explain the tension between theory and experiment in the Tevatron top-quark forward-backward asymmetry. For this reason, it is important to obtain very precise predictions for the pair invariant mass distribution in QCD. In contrast to the case for the total cross section, where results are now known to next-to-next-to-leading order (NNLO) in fixed-order perturbation theory \cite{Baernreuther:2012ws, Czakon:2012pz, Czakon:2013goa}, a full NNLO QCD calculation for the pair invariant mass distribution (or any other distribution) is not yet available. The current state-of-the-art calculation for this distribution combines NLO fixed-order results with soft-gluon resummation at next-to-next-to-leading logarithmic (NNLL) accuracy \cite{Ahrens:2010zv}.  

The main purpose of this paper is to obtain a new set of NNLO QCD corrections to the top-pair invariant mass distribution. In particular, we use the factorization formalism of \cite{Ferroglia:2012ku} to obtain an NNLO soft plus virtual approximation to the differential cross section, valid up to easily quantifiable corrections in powers of $m_t^2/M^2$, with $M$ the top-pair invariant mass. Our results are thus useful for describing highly boosted top-quark production, i.e. the region of phase space where $m_t \ll M$ and such corrections are small. The NNLO corrections obtained here contain all singular terms in the soft limit $z=M^2/\hat{s}\to 1$ (with $\hat{s}$ the partonic center-of-mass energy squared), multiplied by coefficients depending on $m_t$ and the Mandelstam variables. In fact, the coefficients of the logarithmic plus distribution corrections related to soft gluon emission were calculated as an exact function of $m_t$ in \cite{Ahrens:2009uz, Ahrens:2010zv}, so only the delta-function piece needs to be expanded in the limit $m_t \ll M$. Up to corrections to that limit, which quickly become small at larger $M$, our results are equivalent to the NNLO truncation of an NNNLL resummation formula\footnote{More precisely, they provide
the NNLO boundary conditions to the renormalization-group equations used to resum large logarithms.  A full NNNLL resummation requires NNLO  anomalous dimensions which we have not calculated.}, 
and represent the most complete fixed-order result for a differential distribution in top-quark production obtained so far.

After introducing some notation in Section~\ref{sec:notation}, we explain in Section~\ref{sec:softvirtual} how to obtain these NNLO corrections from the factorization formalism of \cite{Ferroglia:2012ku}. Briefly, the building blocks and their interpretations are as follows: \emph{i) Hard functions}: finite contributions from the virtual corrections to the partonic processes $q \bar{q} \to Q \bar{Q}$ and $gg \to Q \bar{Q}$, where $Q$ are massless quarks; \emph{ii) Soft functions}: soft gluon emission corrections to these same partonic processes;  \emph{iii) A heavy-quark fragmentation function:} real and virtual corrections capturing collinear singularities arising in the limit $m_t\to 0$; and \emph{iv) Heavy-flavor matching coefficients:} real and virtual corrections containing $m_t$ dependence induced when matching six-flavor parton distribution functions (PDFs) onto five-flavor ones. Most of these building blocks can be taken directly from previous NNLO calculations: the soft functions were obtained in \cite{Ferroglia:2012uy}, the heavy-quark fragmentation function in \cite{Melnikov:2004bm}, and the heavy-flavor matching coefficients in \cite{Chuvakin:2001ge}. Obtaining the contributions from the hard functions is more subtle. We calculate them here, using the dimensionally regularized NNLO virtual corrections for massless $2 \to 2$ processes from \cite{Anastasiou:2000kg, Anastasiou:2001sv, Anastasiou:2000mv, Glover:2001af, Glover:2001rd} as a starting point. Those results contain IR poles in the dimensional regulator $\epsilon=(4-d)/2$, with $d$ the number of space-time dimensions. We subtract these out using the IR renormalization procedure explained in Section~\ref{sec:virtual}. A consistent implementation of this procedure requires certain color decomposed one-loop corrections up to order $\epsilon^2$, which are not available in the literature and are calculated in this work.

In Section~\ref{sec:pheno} we investigate the phenomenological impact
of our results.  We begin by constructing an improved NNLO
approximation for the invariant mass distribution, which adds the new
 delta-function terms in the small-mass limit calculated in this work to the threshold
enhanced logarithmic plus distribution terms, exact in $m_t$,
determined by NNLL soft-gluon resummation in \cite{Ahrens:2009uz,
  Ahrens:2010zv}.  We then address two main points.  First, we compare
the improved approximation with previous ones based on the logarithmic
terms alone.  We find that the differences between them are rather
small, with the improved approximation providing moderate enhancements
of the differential cross section which are slightly more significant
at lower values of invariant mass than at higher ones. We also explore
the implications of this on approximations for the total inclusive
cross section. Second, we compare the invariant mass distribution
calculated using the newly obtained NNLO approximation with that from
NNLL soft-gluon resummation \cite{Ahrens:2010zv}.  In this case we
find that at low invariant mass the fixed-order predictions are
reliable and higher-order corrections included in the NNLL resummed
results are small. On the other hand, at high invariant mass, the
perturbative convergence is rather poor and the resummation of soft
gluon effects is mandatory.  We comment further on these findings when
concluding in Section~\ref{sec:conclusions}. Some details and results
related to the IR subtraction procedure are given in the appendix.

\section{Notation}
\label{sec:notation}

We study the top-quark pair-production process
\begin{align}
  N_1 (P_1) + N_2(P_2) \rightarrow t(p_3) + \bar{t}(p_4) +X(p_X) \, , \label{eq:hadpr}
\end{align}
where $N_1$ and $N_2$ indicate incoming hadrons and $X$ indicates a hadronic final state including all of the QCD emissions with the exception of the top-antitop pair. The top and the antitop quarks are considered as on-shell particles. At lowest order in QCD, two partonic channels contribute to the process in Eq.~(\ref{eq:hadpr}): the quark annihilation channel
\begin{align}
  q(p_1) + \bar{q}(p_2) \rightarrow t(p_3) + \bar{t}(p_4) \, ,
\end{align}
and the gluon fusion channel
\begin{align}
  g(p_1) + g(p_2) \rightarrow t(p_3) + \bar{t}(p_4) \, .
\end{align}
The momenta of the incoming partons are related to the momenta of the incoming hadrons through $p_i =x_i P_i$
($i = 1,2$). The Mandelstam invariants relevant in this process are
\begin{align}
  s = (P_1 + P_2)^2  \, ,\quad \hat{s} = (p_1+p_2)^2 \, , \quad M^2 = (p_3+p_4)^2 \, , \nonumber
  \\
  t_1 = (p_1-p_3)^2 -m_t^2 \, , \quad u_1 = (p_2-p_3)^2 -m_t^2 \, .
\end{align}
In our formulas we make use of the following dimensionless parameters
\begin{align}
  \tau = \frac{M^2}{s} \, , \quad  z = \frac{M^2}{\hat{s}} \, , \quad \beta_t = \sqrt{1- \frac{4 m_t^2}{M^2}} \, .
\end{align}
The soft gluon limit is then defined by $z\to 1$. In the soft limit the variables $t_1$ and $u_1$ are related to the scattering angle $\theta$ in the partonic center-of-mass frame through the equations
\begin{align}
  t_1 = -\frac{M^2}{2} \left(1 - \beta_t  \cos \theta\right) \, , \quad u_1 = -\frac{M^2}{2} \left(1 + \beta_t  \cos \theta\right) \, .
\end{align}
Furthermore, in the soft limit the Mandelstam invariants satisfy the relation $M^2 + t_1 +u_1 = 0$.

We shall study the distribution differential with respect to the pair invariant mass and the scattering angle:
\begin{align}
  \frac{d^2\sigma}{dMd\cos\theta} & = \frac{8\pi\beta_t}{3sM} \int_\tau^1 \frac{dz}{z}  \, \ff_{ij} \left( \frac{\tau}{z}, \mu_f \right) C_{ij}(z,M,m_t,\cos\theta,\mu_f) \, . \label{eq:diffM}
\end{align}
The indices $ij$ in Eq.~(\ref{eq:diffM}) indicate the partonic channel. (Throughout this paper, the renormalization scale $\mu_r$  and the factorization scale $\mu_f$ are chosen equal to each other.)
In the soft limit, the only contributing channels are the ones already present at lowest order, and therefore $ij \in \{q \bar{q}, gg\}$. The functions $\ff$ are process-independent partonic luminosities,  defined as convolutions of PDFs:
\begin{align}
  \label{eq:ffdef}
  \ff_{ij}(y,\mu_f) = f_{i/N_1}(y,\mu_f) \otimes f_{j/N_2}(y,\mu_f) \equiv \int_y^1 \frac{dx}{x}  f_{i/N_1} \left( \frac{y}{x}, \mu_f \right) f_{j/N_2}(x,\mu_f)  \, .
\end{align}

We are interested in the NNLO corrections to the perturbative hard-scattering kernels $C_{ij}$ in Eq.~(\ref{eq:diffM}). It will be convenient to discuss these corrections at the level of Laplace-transformed coefficients. We define the Laplace-transformed coefficients as
\begin{align}
  \label{eq:C2Ldef}
  \tilde{c}_{ij}({\mathcal N},M,m_t,\cos\theta,\mu_f) &= \int_0^\infty d\xi \, e^{-\xi {\mathcal N}} \, C_{ij}(z,M,m_t,\cos\theta,\mu_f) \, ,
\end{align}  
where\footnote{ The Laplace transform in Eq.~(\ref{eq:C2Ldef}) is identical to the one introduced in \cite{Ahrens:2010zv} and in  \cite{Ferroglia:2012ku}; the definition of the integration variable is $\xi \equiv 2 E_g/M$, where $E_g = M (1-z)/(2 \sqrt{z})$ is the  energy available for the emission of  final state radiation in addition to the top pair. Studies of Drell-Yan scattering\cite{Xu}, Higgs production \cite{Higgs1,Higgs2} and top pair production \cite{Ahrens:2010zv} near threshold showed
that by keeping the exact expression of $E_g$ in the SCET analysis one can reproduce a set of logarithmic power corrections involving $\ln z /(1 - z)$,
which are indeed present in the analytic results for the fixed-order expansions of the hard scattering
kernels. (See also the analogous discussion in \cite{Ahrens:2011mw} for the 1PI kinematics case.)} $\xi=(1-z)/\sqrt{z}$, and denote their perturbative expansions in terms of $\alpha_s$ with five active flavors as 
\begin{align}
  \tilde{c}_{ij}({\mathcal N},M,m_t,\cos\theta,\mu) &= \alpha_s^2 \Biggl[ \tilde{c}^{(0)}_{ij}({\mathcal N},M,m_t,\cos\theta,\mu) + \left( \frac{\alpha_s}{4 \pi} \right) \tilde{c}_{ij}^{(1)}({\mathcal N},M,m_t,\cos\theta,\mu)  \nonumber
  \\
  &\hspace{3em} + \left( \frac{\alpha_s}{4 \pi} \right)^2 \tilde{c}_{ij}^{(2)}({\mathcal N},M,m_t,\cos\theta,\mu) + \mathcal{O}(\alpha_s^3) \Biggr] \, . \label{eq:Lap}
\end{align}
In the soft limit (${\mathcal N}\to \infty)$ the NNLO coefficients have the explicit form
\begin{align} 
  \label{eq:C2L}
  \tilde{c}_{ij}^{(2)}({\mathcal N},M,m_t,\cos\theta,\mu) &= \sum_{n=0}^4 \tilde{c}_{ij}^{(2,n)}(M,m_t,\cos\theta) \ln^n{\frac{M^2}{\bar{{\mathcal N}}^2 \mu^2}}  + {\mathcal O} \left( \frac{1}{{\mathcal N}}\right) \, ,
\end{align}
where $\bar{{\mathcal N}}={\mathcal N}e^{\gamma_E}$. The coefficients proportional to powers of $L \equiv \ln\left( M^2 / (\bar{{\mathcal N}}^2 \mu^2)\right)$ are determined by NNLL soft-gluon resummation and were obtained in \cite{Ahrens:2009uz, Ahrens:2010zv}. The $L$-independent piece $\tilde{c}^{(2,0)}$ is formally of NNNLL order and only its $\mu$-dependence is known. To calculate this coefficient as an exact function of $m_t$, one would need a full soft plus virtual approximation, i.e., the NNLO virtual corrections along with contributions from NNLO real emission in the soft limit. 
This is a very difficult problem, which would require a lengthy numerical calculation employing the techniques recently applied to the evaluation of the total cross section \cite{Baernreuther:2012ws, Czakon:2012pz, Czakon:2013goa}.
However, we will show in the next section how to use results from \cite{Ferroglia:2012ku} to obtain the first term of $\tilde{c}^{(2,0)}$ in
an expansion around the $m_t \to 0$ limit.\footnote{We define the small-mass limit $m_t\to 0$ as $m_t^2\ll \hat{s},t_1,u_1$.}  The result receives corrections in positive powers of $m_t^2/M^2$, which quickly become small at higher values of the invariant mass. It is thus especially useful for describing boosted top production.

The Laplace-space formalism is convenient for explicit calculations and for soft-gluon resummation. However, we also briefly
discuss the structure of the momentum-space results. The NNLO coefficients have the general form
\begin{align}
  \label{eq:C2}
  C_{ij}^{(2)}(z,M,m_t,\cos\theta,\mu) &= D^{ij}_3 \left[ \frac{\ln^3 (1-z)}{1-z} \right]_+ + D^{ij}_2 \left[ \frac{\ln^2(1-z)}{1-z} \right]_+ \nonumber
  \\
  &\quad + D^{ij}_1 \left[ \frac{\ln(1-z)}{1-z}\right]_+ + D^{ij}_0
  \left[\frac{1}{1-z} \right]_+ + C^{ij}_0\,\delta(1-z) + R^{ij}(z) \, .
\end{align}
The coefficients $D_0,\ldots,D_3$ and $C_0$ are functions of the variables $M$, $m_t$, $\cos\theta$ and $\mu$. $D_i$ are determined exactly by NNLL soft-gluon resummation, but $C_0$ and $R(z)$ are not. The calculations presented here determine $C_0$ as an expansion in the small-mass limit. The function $R(z)$ contains terms which are regular in the $z\to 1$ limit and can only be determined through a full NNLO calculation.

\section{Factorization in the double soft-gluon and small-mass limit}
\label{sec:softvirtual}

In \cite{Ferroglia:2012ku} it was shown how, in the double soft and small-mass limit, the hard-scattering kernels factorize into the convolution of several functions:
\begin{align}
  C_{ij}(z,M, m_t, \cos \theta, \mu_f)  &= C^2_D(m_t,\mu_f) \Tr \left[ \bm{H}_{ij}(M,t_1,\mu_f) \, \bm{S}_{ij} \left( \sqrt{\hat{s}} (1-z), t_1, \mu_f \right) \right]  \nonumber
  \\
  &\otimes C_{f\!\!f}^{ij}(z,m_t,\mu_f) \otimes C_{t/t}(z,m_t,\mu_f) \otimes C_{t/t}(z,m_t,\mu_f) \nonumber
  \\
  &\otimes S_D(m_t(1-z),\mu_f) \otimes S_D(m_t(1-z),\mu_f) + {\mathcal O}(1-z) + {\mathcal O}\left(\frac{m_t^2}{M^2}\right) \, .
  \label{eq:facC}
\end{align}
The origin of each of the factors in Eq.~(\ref{eq:facC}) was described in detail in Section 3 of \cite{Ferroglia:2012ku}. Here we simply remind the reader that $\bm{H}$ and $\bm{S}$ are the hard and soft functions for the production of massless top quarks, while $C_D$ and $S_D$ are the collinear and soft-collinear parts of the top-quark fragmentation function, respectively. Finally, the heavy-flavor matching coefficients $C_{f\!\!f}$ and $ C_{t/t} $ are proportional to powers of $n_h=1$ and arise from the fact that PDFs and fragmentation functions are written in terms of $\alpha_s$ with $n_l=5$ active flavors.  

The factorization formula is simpler to discuss in Laplace space, where it becomes a product of the various functions. We define Laplace transforms of the $z$-dependent functions as
\begin{align}
  \label{eq:LaplaceTransforms}
  \tilde{\bm{s}}_{ij} \Biggl( \ln\frac{M^2}{\bar{{\mathcal N}}^2\mu_f^2}, t_1, \mu_f \Biggr) &= \int_0^\infty d\xi \, e^{-\xi {\mathcal N}} \, \bm{S}_{ij}\left(\sqrt{\hat{s}}(1-z),t_1,\mu_f\right) \, , \nonumber
  \\
  \tilde{s}_D \Biggl( \ln\frac{m_t}{\bar{{\mathcal N}}\mu_f},\mu_f \Biggr) &= \int_0^\infty d\xi \, e^{-\xi {\mathcal N}} \, S_D(m_t(1-z),\mu_f) \, , \nonumber
  \\
  \tilde{c}_t^{ij} \Biggl( \ln\frac{1}{\bar{{\mathcal N}}^2}, m_t, \mu_f \Biggr) &= \int_0^\infty d\xi \, e^{-\xi {\mathcal N}} \, C^{ij}_{\fff}(z,m_t,\mu_f) \otimes C_{t/t}(z,m_t,\mu_f) \otimes C_{t/t}(z,m_t,\mu_f) \, .
\end{align}
As above, $\xi=(1-z)/\sqrt{z}$ and $\bar{{\mathcal N}}={\mathcal N}e^{\gamma_E}$. The first argument of the Laplace transforms is more conveniently written in terms of 
\begin{align}
  L\equiv \ln\frac{M^2}{\bar{{\mathcal N}}^2\mu^2}  \, , \quad L' \equiv   \ln\frac{m_t^2}{\bar{{\mathcal N}}^2\mu^2}=L-\ln\frac{M^2}{m_t^2} \, , \quad \text{and} \quad L'' \equiv \ln\frac{1}{\bar{{\mathcal N}}^2} = L-\ln\frac{M^2}{\mu^2} \, .
\end{align}
The factorization formula for the Laplace-transformed functions in Eq.~(\ref{eq:C2Ldef}) then reads
\begin{align}
  \tilde{c}_{ij}({\mathcal N},M,m_t,\cos\theta,\mu) &= C_D^2(m_t,\mu) \, \Tr \left[ \bm{H}_{ij}(M,\cos\theta,\mu) \, \tilde{\bm{s}}_{ij}(L,\cos\theta,\mu) \right] \tilde{s}_D^2(L'/2,\mu) \, \tilde{c}_t^{ij}(L'',\mu) \, .
  \label{eq:cexp}
\end{align}

In order to express the NNLO corrections to the Laplace space coefficients $\tilde{c}_{ij}$ in terms of the component functions we first define expansion coefficients in powers of $\alpha_s$ with five active flavors as (here and in the rest of the section we suppress the subscripts labeling the channel dependence and the arguments of the functions):
\begin{align}
  \bm{H} &= \alpha_s^2 \, \frac{3}{8d_R} \left[ \bm{H}^{(0)} + \left( \frac{\alpha_s}{4\pi} \right) \bm{H}^{(1)} + \left( \frac{\alpha_s}{4\pi} \right)^2 \bm{H}^{(2)}  + {\mathcal O}(\alpha_s^3) \right] , \nonumber
  \\
  \tilde{\bm{s}} &= \tilde{\bm{s}}^{(0)} + \left( \frac{\alpha_s}{4\pi} \right) \tilde{\bm{s}}^{(1)} + \left( \frac{\alpha_s}{4\pi} \right)^2 \tilde{\bm{s}}^{(2)} + {\mathcal O}(\alpha_s^3) \, , \nonumber
  \\
  C_D &= 1 + \left( \frac{\alpha_s}{4\pi} \right) C_D^{(1)} + \left( \frac{\alpha_s}{4\pi}\right)^2 C_D^{(2)} + {\mathcal O}(\alpha_s^3) \, , \nonumber
  \\
 \tilde{s}_D &= 1 + \left( \frac{\alpha_s}{4\pi} \right) \tilde{s}_D^{(1)} + \left( \frac{\alpha_s}{4\pi} \right)^2 \tilde{s}_D^{(2)} + {\mathcal O}(\alpha_s^3) \, , \nonumber
 \\
 \tilde{c}_t &= 1 + \left( \frac{\alpha_s}{4\pi} \right) \tilde{c}_t^{(1)} + \left( \frac{\alpha_s}{4\pi} \right)^2 \tilde{c}_t^{(2)} + {\mathcal O}(\alpha_s^3) \, .
 \label{eq:exp}
\end{align}
The factor $d_R$ is the number of colors $N$ in the quark annihilation channel and $N^2-1$ in the gluon fusion channel. By inserting the expansions in Eqs.~(\ref{eq:exp}) into Eq.~(\ref{eq:cexp}), we then find
\begin{align}
  \tilde{c}^{(0)} &= \frac{3}{8 d_R} \Tr \left[ \bm{H}^{(0)} \tilde{\bm{s}}^{(0)}  \right] , \nonumber
  \\
  \tilde{c}^{(1)} &= \frac{3}{8 d_R} \Tr \left[ \bm{H}^{(1)} \tilde{\bm{s}}^{(0)} + \bm{H}^{(0)} \tilde{\bm{s}}^{(1)} + \left( c_t^{(1)} + 2 C_D^{(1)} + 2 \tilde{s}_D^{(1)} \right) \bm{H}^{(0)} \tilde{\bm{s}}^{(0)} \right] , \nonumber
  \\
  \tilde{c}^{(2)} &=\frac{3}{8 d_R} \Tr \Biggl\{ \bm{H}^{(2)} \tilde{\bm{s}}^{(0)} + \bm{H}^{(0)} \tilde{\bm{s}}^{(2)} + \bm{H}^{(1)} \tilde{\bm{s}}^{(1)} + \left( c_t^{(1)} + 2 C_D^{(1)} + 2 \tilde{s}_D^{(1)} \right) \left( \bm{H}^{(0)} \tilde{\bm{s}}^{(1)} + \bm{H}^{(1)} \tilde{\bm{s}}^{(0)} \right) \nonumber
  \\
  &+ \left[ c_t^{(2)} + 2 C_D^{(2)} + 2 \tilde{s}_D^{(2)} + \left(C_D^{(1)}\right)^2 + \left(\tilde{s}_D^{(1)}\right)^2 + 2 c_t^{(1)} \left(  \tilde{s}_D^{(1)}+ C_D^{(1)} \right) + 4 C_D^{(1)} \tilde{s}_D^{(1)} \right] \bm{H}^{(0)} \tilde{\bm{s}}^{(0)} \Biggr\} \, .
  \label{eq:cts}
\end{align}
All of the terms in the r.h.s. of the Eq.~(\ref{eq:cts}) can be assembled in a straightforward way starting from the results collected in \cite{Ferroglia:2012ku, Ferroglia:2012uy}, except for the term $\Tr \left[ \bm{H}^{(2)} \tilde{\bm{s}}^{(0)} \right]$, which involves the NNLO hard function. In the following section we explain how to extract this missing piece, using as a starting point the NNLO virtual corrections from \cite{Anastasiou:2000kg, Anastasiou:2001sv, Anastasiou:2000mv}.

\section{NNLO  virtual corrections for massless top quarks}
\label{sec:virtual}

In this section we explain our method for extracting the contribution of the NNLO hard function to Eq.~(\ref{eq:cts}). The starting point is the dimensionally regularized NNLO virtual corrections to massless $q\bar{q} \to Q\bar{Q}$ and $gg \to Q\bar{Q}$ scattering, which were evaluated more than a decade ago in \cite{Anastasiou:2000kg, Anastasiou:2000mv, Anastasiou:2001sv}. However, these results themselves are not sufficient: they are UV renormalized but still contain IR poles in the dimensional regulator $\epsilon$. One must supplement them with an IR subtraction procedure, which not only cancels the poles but also adds certain finite contributions. We outline this procedure below, and then give some more details along with explicit results in the appendix.

We first set up some notation related to the color-space formalism of \cite{Catani:1996jh}. This basis-independent notation applies equally well to the gluon fusion and quark-antiquark annihilation channels. Channel-dependent results for a particular color basis are given in the appendix. We thus denote the UV renormalized $gg \to Q\bar{Q}$ or $q\bar{q} \to Q\bar{Q}$ scattering amplitudes by a color-space vector, whose expansion in $\alpha_s$ is
\begin{align}
  \ket{\mathcal{M}(\epsilon,\hat{s},t_1)} = 4\pi\alpha_s \left( \ket{\mathcal{M}_0} + \frac{\alpha_s}{4\pi} \ket{\mathcal{M}_1} + \left(\frac{\alpha_s}{4\pi}\right)^2 \ket{\mathcal{M}_2} + \cdots \right) .
\end{align}
Here and below, the arguments of the expansion coefficients are suppressed. The squared matrix element, summed over colors and spins, is denoted by the inner product $\braket{ \mathcal{M} | \mathcal{M} }$. We define the perturbative expansion of this quantity as
\begin{align}
  \braket{\mathcal{M}(\epsilon,\hat{s},t_1) | \mathcal{M}(\epsilon,\hat{s},t_1) } = 16\pi^2\alpha_s^2 \left[ \mathcal{V}^{(0)} + \frac{\alpha_s}{4\pi} \mathcal{V}^{(1)} + \left( \frac{\alpha_s}{4\pi} \right)^2 \mathcal{V}^{(2)} + \cdots \right] .
\end{align}
It is convenient to split up the NNLO corrections as
\begin{align}
  {\mathcal V}^{(2)}={\mathcal V}^{(2\times 0)} + {\mathcal V}^{(1\times 1)} \, ,
\end{align}
where ${\mathcal V}^{(2\times 0)}$ denotes the interference of the two-loop diagrams with the tree-level amplitude, and ${\mathcal V}^{(1\times 1)}$ the square of one-loop diagrams. In the color-space notation, we then have
\begin{align}
  {\mathcal V}^{(2\times 0)} = \braket{ {\mathcal M}_0 | {\mathcal M}_2 } + \braket{ {\mathcal M}_2 | {\mathcal M}_0 } \, , \quad \text{and} \quad {\mathcal V}^{(1\times 1)} =  \braket{ {\mathcal M}_1 | {\mathcal M}_1 } \, .
\end{align}
Results for these terms can be extracted from the literature. In the gluon fusion channel, we have ${\mathcal V}^{(2 \times 0)} = 4 \, {\mathcal C}^{8(2 \times 0)}$ and ${\mathcal V}^{(1 \times 1)} = 4 \, {\mathcal C}^{8(1 \times 1)}$, where ${\mathcal C}^{8(2 \times 0)}$ and  ${\mathcal C}^{8(1 \times 1)}$ can be read off from Eq.~(3.1) and Eq.~(4.1) of \cite{Anastasiou:2001sv}, respectively. For the quark annihilation channel, ${\mathcal V}^{(2 \times 0)} = 4 \, {\mathcal A}^{8(2 \times 0)}$ and ${\mathcal V}^{(1 \times 1)} = 4 \, {\mathcal A}^{8(1 \times 1)}$, where ${\mathcal A}^{8(2 \times 0)}$ and ${\mathcal A}^{8(1 \times 1)}$ can be taken from Eq.~(4.2) of \cite{Anastasiou:2000kg} and Eq.~(4.1) of \cite{Anastasiou:2000mv}, respectively.\footnote{The factors of four arise because we expand the squared amplitude in terms of $\alpha_s/(4\pi)$ instead of $\alpha_s/(2\pi)$ as in \cite{Anastasiou:2000kg, Anastasiou:2000mv, Anastasiou:2001sv}. We have used an  electronic form of the results for the gluon-fusion one-loop squared amplitudes given to us by Nigel Glover rather than extracting them from \cite{Anastasiou:2001sv}.}

The scattering amplitudes contain IR poles in the dimensional regulator $\epsilon$. These can be subtracted by the renormalization procedure described in \cite{Becher:2009cu, Becher:2009qa} (see also \cite{Catani:1998bh}), which amounts to evaluating the following equation:
\begin{align}
 \ket{{\mathcal M}^{\text{ren}}(\hat{s},t_1,\mu)} &= \lim_{\epsilon \to 0} \bm{Z}^{-1}(\epsilon,\hat{s},t_1,\mu) \, \ket{{\mathcal M}(\epsilon,\hat{s},t_1)} \, ,  \nonumber
 \\
 &= \lim_{\epsilon \to 0} \left( \bm{1} + \frac{\alpha_s}{4 \pi} \bm{Z}_1 +\left( \frac{\alpha_s}{4 \pi} \right)^2 \bm{Z}_2 + \cdots \right)^{-1} \ket{{\mathcal M}(\epsilon,\hat{s},t_1)} \, .
 \label{eq:renM}
\end{align}
The superscript ``ren'' in the l.h.s. of Eq.~(\ref{eq:renM}) indicates that the IR renormalized amplitude is completely free from poles in $\epsilon$. In particular, the poles in the bare amplitudes $\ket{{\mathcal M}}$ are canceled by those in the multiplicative renormalization factor $\bm{Z}$. This renormalization factor is a matrix in color space, and can be calculated starting from the general formula presented in \cite{Becher:2009cu, Becher:2009qa}.

The square of the finite, renormalized amplitudes can be used to calculate the contributions $\Tr \left[ \bm{H}^{(2)} \tilde{\bm{s}}^{(0)} \right]$. By evaluating $\braket{ {\mathcal M}^{\text{ren}} | {\mathcal M}^{\text{ren}} }$ using Eq.~(\ref{eq:renM}) and extracting the NNLO corrections, one finds certain combinations which are free of IR poles. For the two-loop corrections
\begin{align}
  \braket{ {\mathcal M}_0 | {\mathcal M}_2 } - \braket{ {\mathcal M}_0 | \bm{Z}_1 | {\mathcal M}_1 } + \braket{ {\mathcal M}_0 | \bm{Z}^2_1 | {\mathcal M}_0 } - \braket{ {\mathcal M}_0 | \bm{Z}_2 | {\mathcal M}_0 } &= \text{finite} \, ,
 \label{eq:fin1}
\end{align}
while for the one-loop squared terms 
\begin{align}
  \braket{ {\mathcal M}_1 | {\mathcal M}_1 } - \braket{ {\mathcal M}_1 | \bm{Z}_1 | {\mathcal M}_0 } - \braket{ {\mathcal M}_0 | \bm{Z}^\dagger_1 | {\mathcal M}_1 } + \braket{ {\mathcal M}_0 | \bm{Z}^\dagger_1 \bm{Z}_1 | {\mathcal M}_0 } & = \text{finite} \, .
\end{align}
This leads us to define the following IR counterterms 
\begin{align}
  {\mathcal K}^{(2 \times 0)} = 2 \mathrm{Re} \left[ \braket{ {\mathcal M}_0 | \bm{Z}_1 | {\mathcal M}_1 } - \braket{ {\mathcal M}_0 | \bm{Z}^2_1 | {\mathcal M}_0 } + \braket{ {\mathcal M}_0 | \bm{Z}_2 | {\mathcal M}_0 } \right] ,
  \label{eq:ct02}
\end{align}
and
\begin{align}
  {\mathcal K}^{(1 \times 1)} = \braket{ {\mathcal M}_1 | \bm{Z}_1 | {\mathcal M}_0 } + \braket{ {\mathcal M}_0 | \bm{Z}^\dagger_1 | {\mathcal M}_1 } - \braket{ {\mathcal M}_0 | \bm{Z}^\dagger_1 \bm{Z}_1 | {\mathcal M}_0 } \, .
  \label{eq:ct11}
\end{align}
The final result for the NNLO corrections entering the factorization formula is then
\begin{align}
  \label{eq:matching}
  \Tr \left[ \bm{H}^{(2)} \tilde{\bm{s}}^{(0)} \right] &= \frac{1}{4d_R} \left( {\mathcal V}^{(2)} - {\mathcal K}^{(1 \times 1)} - {\mathcal K}^{(2 \times 0)} \right) \, .
  \\
  &\equiv N^3 A + N B +\frac{1}{N} C+ \frac{1}{N^3} D+ n_l  N^2 E + n_l F + \frac{n_l^2}{N^2} G+ n_l^2 N H  + \frac{n_l^2}{N} I  \,.
\end{align}
In the last line of Eq.~(\ref{eq:matching}) we made the color structure of $\mbox{Tr} \left[ \bm{H}^{(2)}
  \tilde{\bm{s}}^{(0)}\right]$ explicit; the coefficients $A, \cdots, I$ are channel dependent ($I=0$ in the gluon fusion channel) and are functions of $M, t_1$ and $\mu$.
Eq.~(\ref{eq:matching}) follows from the definition of the hard and soft function matrix elements in terms of the renormalized amplitudes and color basis vectors \cite{Ahrens:2010zv}
\begin{align}
  H_{IJ}^{(2)} &= \frac{1}{4} \frac{1}{\braket{c_I|c_I} \, \braket{c_J|c_J}} \Bigl[ \braket{ c_I | {\mathcal M}^{\text{ren}}_0 } \braket{ {\mathcal M}_2^{\text{ren}} | c_J } + \braket{ c_I | {\mathcal M}^{\text{ren}}_2 } \braket{ {\mathcal M}_0^{\text{ren}} | c_J } + \braket{ c_I | {\mathcal M}^{\text{ren}}_1 } \braket{ {\mathcal M}_1^{\text{ren}} | c_J } \Bigr] ,
\end{align}
and $\tilde{s}_{IJ}^{(0)} = \langle c_I| c_J \rangle /d_R$, where $\ket{c_I}$ are color basis vectors whose definition is given in the appendix. The explicit expressions for $\Tr \left[ \bm{H}^{(2)} \tilde{\bm{s}}^{(0)} \right]$ in the two production channels are rather lengthy, and are provided in electronic form with the arXiv submission of this paper.

The most difficult part of this calculation is of course evaluating the NNLO corrections ${\mathcal V}^{(2)}$, which we have taken from \cite{Anastasiou:2000kg, Anastasiou:2000mv, Anastasiou:2001sv}. However, equally important are the IR counterterms $\mathcal{K}$, which are not available in the literature. We have thus calculated them from scratch. Some of the details are discussed in the appendix, where we give results for all of the ingredients needed in our calculation. An important element, obtained here for the first time, is a certain set of color decomposed one-loop amplitudes to order $\epsilon^2$.

While our results are new, we are still able to perform three important consistency checks. The first is that the combination shown in Eq.~(\ref{eq:matching}) is indeed finite in the limit $\epsilon \to 0$. The second is that the $\mu$-dependent terms in the NNLO function are in agreement with those derived using renormalization-group equations presented in \cite{Ferroglia:2012ku}.  The third, explained in the appendix, involves a comparison with two-loop corrections obtained in \cite{Czakon:2007ej, Czakon:2007wk}.

\section{Predictions for the pair invariant mass distribution at the LHC}
\label{sec:pheno}

In this section we explore the impact of our results on the pair
invariant mass distribution at the LHC.  We compare numerical results
within different perturbative approximations, and then make some
general statements concerning the importance of higher-order
corrections, also at the level of the total cross section.

We begin by introducing three different approximations to the NNLO corrections
to the pair invariant mass distribution.  We define these approximations
at the level of the Laplace-transformed coefficients in Eq.~(\ref{eq:C2L}).
In each case, we use the results from \cite{Ahrens:2010zv} for 
the logarithmic coefficients $\tilde{c}_{ij}^{(2,n)}$, with $n=1,2,3,4$.
These are determined from NNLL soft-gluon resummation and are exact in
$m_t$.  We then add to these one of the following three approximations
for the non-logarithmic coefficient $\tilde{c}_{ij}^{(2,0)}:$
\begin{enumerate}
\item[A.] use no information, i.e. $\tilde{c}_{ij}^{(2,0)}=0$; 
\item[B.] use the information from the NNLO fragmentation function,
  plus the $n_h$ terms arising from $\alpha_s$-decoupling and the
  heavy-flavor coefficients, thereby including all terms enhanced by
  (up to two) powers of $\ln m_t/M$ for $\mu_f\sim M$;
\item[C.] use  the $m_t\to 0$ limit of $\tilde{c}_{ij}^{(2,0)}$.
\end{enumerate}
Approximations A and B were considered in \cite{Ferroglia:2012ku}.
Approximation C is the full virtual plus soft approximation 
in the small-mass limit, made possible by the
results presented here.  These approximations are progressively more 
complete.  Approximation A contains only the terms determined by NNLL
soft-gluon resummation for arbitrary $m_t$, and approximation B adds to these the logarithmic 
terms determined by NNLL resummation in the double soft and small-mass limit.  Finally, 
Approximation C contains information determined  from terms which are formally part of the expansion of  NNNLL resummation formulas in the small mass limit.  In what follows, we will compare the
higher-order corrections within these NNLO approximations both with
each other and with those obtained from NNLL soft-gluon resummation.

Our main results are gathered in Tables~\ref{tab:7} and~\ref{tab:14}.
These show numerical values of the invariant mass distribution at the
LHC with two different collider energies, using the inputs described
in the captions.  In each table, the first row shows the LO
distribution and the second row the NLO correction using the leading
terms in the $z\to 1$ limit as evaluated in \cite{Ahrens:2010zv}.  The
next three rows show corrections to the NLO distribution in the $z\to
1$ limit (i.e. the correction to the sum of the first two rows)
obtained using NNLO approximations A--C above, and the final row the
analogous correction but using NNLL soft-gluon resummation as
implemented in \cite{Ahrens:2010zv}.

We note that the new terms contained in approximation C produce roughly a 20--40\%
enhancement to the logarithmic terms in approximation A, and roughly a
10--30\% enhancement to the terms contained in approximation B, the
effect being larger at smaller invariant mass or higher collider
energy. A study of the NLO corrections within the analogous
approximations was carried out for the LHC with $\sqrt{s}=7$~TeV in
\cite{Ferroglia:2012ku}; the numbers provided in Table~1 of that work
show that the extra terms in approximation C are roughly twice as
important at NLO as at NNLO.  This is expected, since at NLO the
Laplace-space coefficient contains only two powers of the threshold
logarithm while at NNLO it contains four.  It is also expected that
the logarithmic terms are larger at high invariant mass, since larger
$M$ is characterized by larger average value of $z$.  By the same
token, the naive expectation is that the non-singular terms in the $z\to 1$ limit are expected to be
smaller at larger $M$. To the extent this is true, approximation C
should be very close to the exact NNLO result at high values of $M$,
since the subleading terms in $m_t/M$ quickly become small as the
invariant mass is increased.

Of course, the size of power corrections
to the soft limit as a function of $M$ will only be known for sure
once the full NNLO calculation of the differential cross section is
completed. However, we can get a rough idea of how well the approximation
works in  the low-invariant mass region by studying the total inclusive
cross section.  We do so at the end of this section. Another option is to study how 
well the soft plus virtual approximation works at NLO. In Tables~\ref{tab:7nlo}
and~\ref{tab:14nlo} we compare the exact NLO correction obtained from MCFM \cite{Campbell:2000bg} with
the leading terms in the $z\to 1$ limit captured by the soft plus virtual approximation.  
We show results for bins of invariant mass centered around the three values used so far, and have 
used the same input as in Tables~\ref{tab:7} and Tables~\ref{tab:14}.  Evidently, the soft plus 
virtual approximation works quite well at NLO.  

\begin{table}[t!]
  \centering
  \begin{tabular}{|c|c|c|c|}
    \hline
    & $M=500$~GeV & $M=1500$~GeV & $M=3000$~GeV
    \\ \hline\hline
    LO & $1.89 \times 10^{-1}$ & $1.65 \times 10^{-4}$ & $9.40 \times 10^{-8}$
    \\ \hline\hline
    NLO corr. ($z\to 1$) & $1.54 \times 10^{-1}$ & $1.86 \times 10^{-4}$ & $1.20 \times 10^{-7}$
    \\ \hline\hline
    NNLO corr. (approx. A) & $5.67 \times 10^{-2}$ & $1.22 \times 10^{-4}$ & $1.11 \times 10^{-7}$
    \\ \hline
    NNLO corr. (approx. B) & $6.35 \times 10^{-2}$ & $1.40 \times 10^{-4}$ & $1.26 \times 10^{-7}$
    \\ \hline
    NNLO corr. (approx. C) & $7.31 \times 10^{-2}$ & $1.52 \times 10^{-4}$ & $1.33 \times 10^{-7}$
    \\ \hline \hline
    NNLL corr. & $8.40 \times 10^{-2}$ & $2.69 \times 10^{-4}$ & $3.97 \times 10^{-7}$
    \\ \hline
  \end{tabular}
  \caption{\label{tab:7} The LO differential cross section  $d\sigma/dM$ (in pb/GeV) at the LHC with $\sqrt{s}=7$~TeV, along with higher-order QCD corrections obtained as described in the text.  We use $m_t=172.5$~GeV, $\mu_f= \mu_r=M$ and MSTW2008NNLO PDFs \cite{Martin:2009iq}. Approximations A, B and C differ only in their treatment of $\tilde{c}_{ij}^{(2,0)}$, as explained in the text. The numbers in the table include the corresponding power of $\alpha_s$, so that the differential cross section at approximate NNLO can be obtained by summing the numbers in the lines labeled LO, NLO corr. ($z \to 1$), and  NNLO corr. (approx i) (i=A,B,C). Similarly, the differential cross section at NNLL accuracy can be obtained by summing the LO, NLO corr. ($z \to 1$) and NNLL corr. lines.}
\end{table}


\begin{table}[t!]
  \centering
  \begin{tabular}{|c|c|c|c|}
    \hline
    & $M=500$~GeV & $M=1500$~GeV & $M=3000$~GeV
    \\ \hline\hline
    LO & 1.11 & $3.50 \times 10^{-3}$ & $2.04 \times 10^{-5}$
    \\ \hline\hline
    NLO corr. ($z \to 1$) & $8.58 \times 10^{-1}$ & $3.74 \times 10^{-3}$ & $2.51 \times 10^{-5}$
    \\ \hline\hline
   NNLO corr.  (approx. A) & $2.64 \times 10^{-1}$ & $2.00 \times 10^{-3}$ & $1.77 \times 10^{-5}$
    \\ \hline
  NNLO corr.   (approx. B) & $3.05 \times 10^{-1}$ & $2.40 \times 10^{-3}$ & $2.11 \times 10^{-5}$
    \\ \hline
   NNLO corr.  (approx. C) & $3.65 \times 10^{-1}$ & $2.67 \times 10^{-3}$ & $2.31 \times 10^{-5}$
    \\ \hline \hline
    NNLL corr. & $3.72 \times 10^{-1}$ & $3.79 \times 10^{-3}$ & $4.42 \times 10^{-5}$
    \\ \hline
  \end{tabular}
  \caption{\label{tab:14} Same as Table~\ref{tab:7}, but with $\sqrt{s}=14$~TeV.}
\end{table}


\begin{table}[t!]
  \centering
\begin{tabular}{|c||c|c|c|}
\hline bin [GeV] &  LO [pb/GeV]& NLO corr. [pb/GeV]  & NLO corr. $z \to 1$ [pb/GeV]   \\ 
\hline
\hline 400--600 & $0.206$ & $0.158$ & $0.163$  \\ 
\hline  1400--1600 & $1.74 \times10^{-4}$  & $1.72  \times10^{-4}$&  $1.96  \times10^{-4}$\\ 
\hline 2900--3100 & $9.76 \times 10^{-8}$ & $1.21 \times 10^{-7}$  & $1.26 \times 10^{-7}$ \\ 
\hline 
\end{tabular}  \caption{\label{tab:7nlo} Comparison between the NLO exact and NLO leading corrections in 
the $z\to1$ limit at the  LHC with $\sqrt{s}=7$~TeV. The numbers refer to different bins in  the pair invariant mass.}
\end{table}


\begin{table}[t!]
  \centering
\begin{tabular}{|c||c|c|c|}
\hline bin [GeV] &  LO [pb/GeV]& NLO corr. [pb/GeV]  & NLO corr. $z \to 1$ [pb/GeV]   \\ 
\hline
\hline 400--600 & $1.16$ & $0.86$ & $0.88$  \\ 
\hline  1400--1600 & $3.61 \times10^{-3}$  & $ 3.70 \times10^{-3}$&  $ 3.87 \times10^{-3}$\\ 
\hline 2900--3100 & $2.07 \times 10^{-5}$ & $ 2.18 \times 10^{-5}$  & $ 2.55 \times 10^{-5}$ \\ 
\hline 
\end{tabular}  \caption{\label{tab:14nlo} Same as Table~\ref{tab:7nlo}, but with $\sqrt{s}=14$~TeV.}
\end{table}

\begin{figure}[t!]
   \centering
   \includegraphics[width=0.65\textwidth]{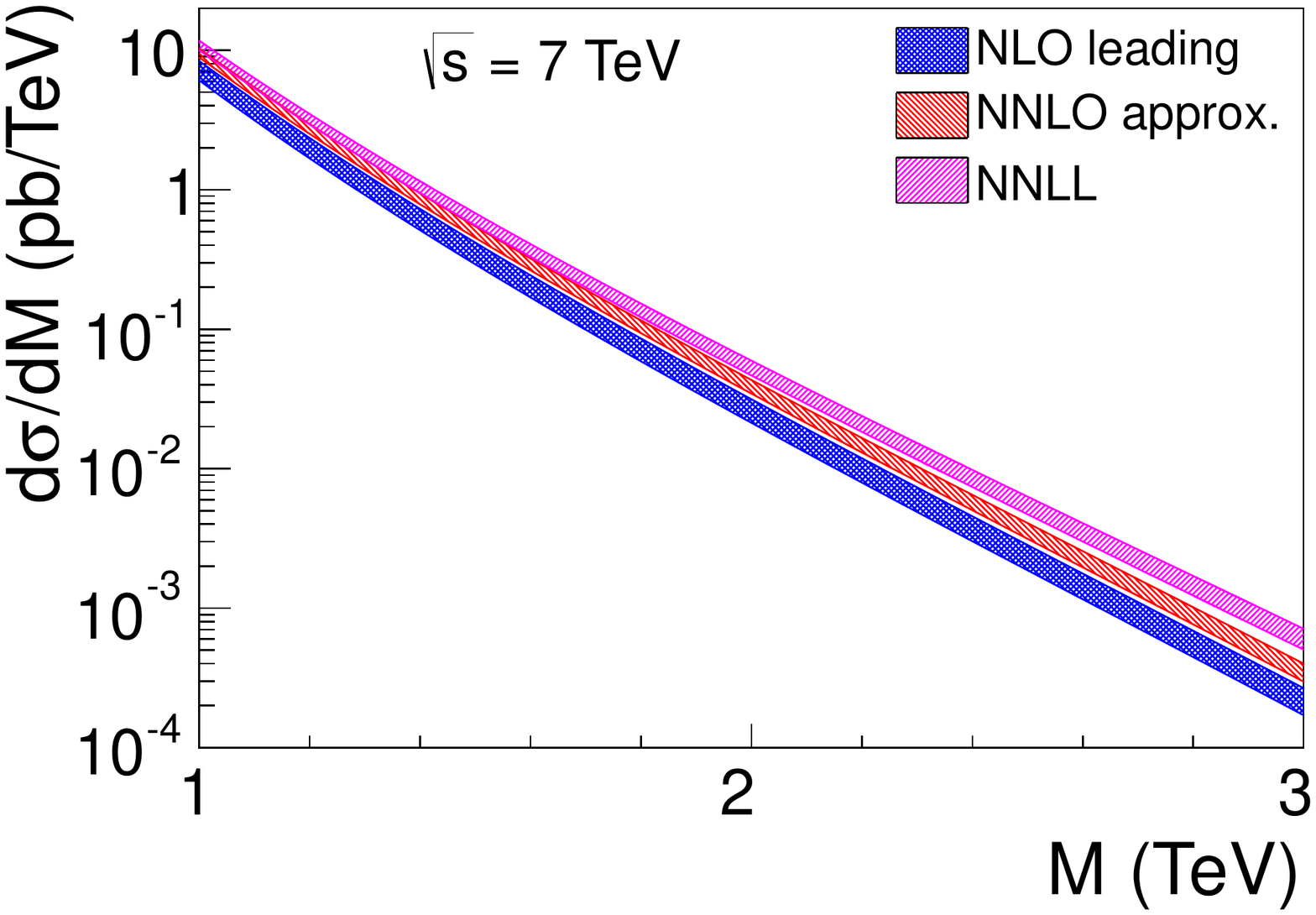}
   \includegraphics[width=0.65\textwidth]{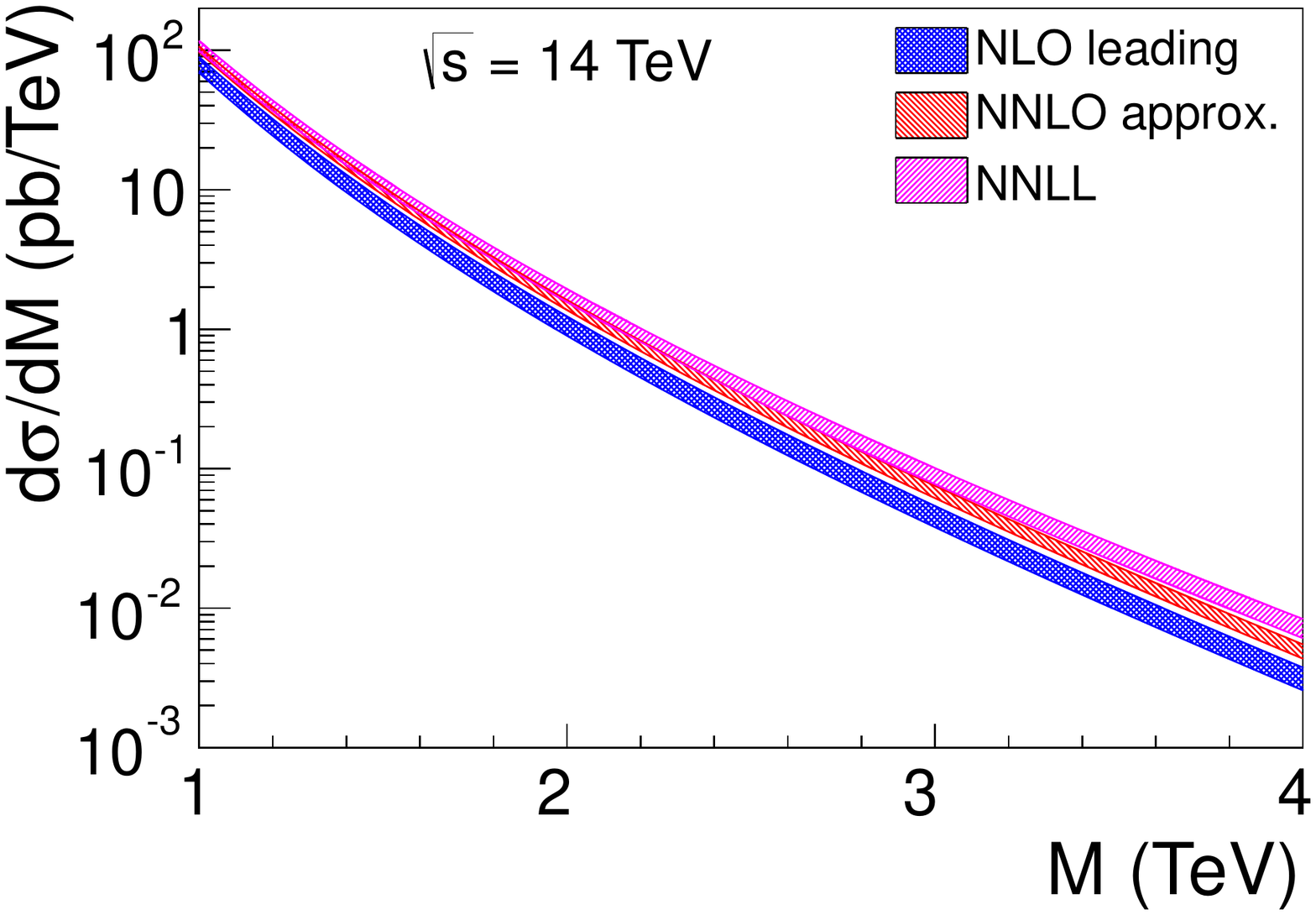}
   \caption{Invariant mass distribution at the LHC with $\sqrt{s}=7$~TeV (upper panel) and 14~TeV (lower panel).}
   \label{fig:M}
 \end{figure}

We next comment on the convergence of the perturbative series and the
importance of soft-gluon resummation.  We first examine the case of
$M=500$~GeV, which we consider representative of relatively low
invariant mass.  In that case, the numbers in Tables~\ref{tab:7}
and~\ref{tab:14} show that the NNLO corrections are rather mild
compared to the NLO ones, with the NNLL corrections providing only a
slight further enhancement.  The perturbative series converges well
and fixed-order perturbation theory is reliable.  For the high
invariant-mass region the situation is quite different.  For instance,
at $M=3000$~GeV, the NNLO corrections are roughly the same size as the
NLO ones, and the NNLL corrections are larger still.  Fixed-order
perturbation theory thus breaks down, and the fact that the bulk of
the NNLO correction is provided by the combination of soft and
small-mass logarithmic terms included in approximation B gives a
strong motivation to use resummation in that region.  We further
illustrate these points through the results in Figure~\ref{fig:M}. There we compare the
invariant mass distribution using the leading terms in the $z\to 1$
limit at NLO (labeled NLO leading), the leading terms in the $z\to1$
limit with the corrections from approximation C added on (labeled NNLO
approx.), and the NNLL calculation from \cite{Ahrens:2010zv}.  The
bands reflect scale uncertainties obtained by varying the
factorization scale in the range $M/2<\mu_f<2M$ (the NNLL result
depends in addition on hard and soft matching scales, which are varied
as in \cite{Ahrens:2010zv}). Obviously, the conclusions drawn above
are not changed once scale uncertainties are taken into account.

The overall picture that emerges from our numerical study above is that
while at lower values of invariant mass soft-gluon resummation adds
only small enhancements to the differential cross section, at higher
values of invariant mass it can have quite a large effect and should
not be neglected. We comment on possibilities for further studies in
the conclusions.

\begin{table}[t!]
  \centering
  \begin{tabular}{|c|c|c|c|}
    \hline
    $\sigma$ [pb] & Tevatron & LHC $7$ TeV & LHC $14$ TeV
    \\ \hline
    NNLO approx. PIM & $6.573^{+0.050}_{-0.395}$ & $153.8^{+8.1}_{-8.2}$ & $855.3^{+46.4}_{-42.2}$
    \\ \hline
    NNLO approx. C & $6.683^{+0.179}_{-0.372}$& $156.7^{+10.0}_{-7.0}$ & $873.2^{+60.8}_{-34.6}$
    \\ \hline
    NNLO exact \cite{Czakon:2013goa} & $7.009^{+0.259}_{-0.374}$ &  $167.0^{+6.7}_{-10.7}$ & $933.0^{+31.8} _{-51.0}$
    \\ \hline 
  \end{tabular} 
  \caption{\label{tab:3} Total cross section with scale uncertainty estimated as described in the text, using $m_t =173.3$~GeV and MSTW2008NNLO PDFs.}
\end{table}

We end this section by discussing in more detail the NNLO corrections
to the total cross section.  The exact NNLO results were recently
obtained in \cite{Baernreuther:2012ws, Czakon:2012pz, Czakon:2013goa}
and lead to total cross section predictions which are higher than the
ones obtained by integrating NNLO approximations to the invariant mass
distribution obtained from NNLL soft-gluon resummation. Since the
total cross section receives its dominant contributions from low
values of invariant mass, comparing the exact NNLO cross section with
approximate ones gives some idea of the agreement between them in
that region of phase space, although obviously any direct information
about the shape of the distribution is lost in the integration
process.  In Table~\ref{tab:3} we compare the total cross section
obtained by integrating different NNLO approximations to the pair
invariant mass distribution to the exact calculation of the total
cross section at NNLO \cite{Czakon:2013goa}. The central value
corresponds to $\mu_f=m_t$, and scale uncertainties are estimated by
evaluating the cross section at $\mu_f=m_t/2$ (the upper numbers) and
$\mu_f=2m_t$ (the lower numbers).  The row labeled ``NNLO
approx. PIM'' in Table~\ref{tab:3} corresponds to the approximation
used for the invariant mass distribution in computing the total cross 
section in \cite{Ahrens:2011mw, Ahrens:2011px}; 
 i.~e.,  only part of the scale dependent terms in $\tilde{c}_{ij}^{(2,0)}$ are included, together with a set of terms which are regular in the $z \to 1$ limit and are obtained by keeping the exact form of the soft emission energy in the SCET formalism, as explained in detail in \cite{Ahrens:2011mw}. 
 Approximation C is as described above. All entries in Table~\ref{tab:3} include the full NLO corrections.  Table~\ref{tab:3} shows that
the inclusion of the corrections in approximation C helps decrease the
gap between the exact NNLO result and the approximate NNLO
calculations. In particular, at the LHC the range of values determined
by perturbative uncertainty in approximation C has a sizable overlap
with the uncertainty range of the complete NNLO prediction. 
This effect could not be predicted before carrying out the calculation of the soft plus virtual approximation presented in this work. However, it can be understood a posteriori on the basis of the fact that the corrections in approximation C increase the differential cross section in a region of invariant mass values  which, upon integration, provides a large contribution to the total cross section. This fact can be seen by looking at Tables~\ref{tab:7} and  \ref{tab:14} and at Figure \ref{fig:M}.
 The
remaining difference between approximation C and the exact result for the total cross section at NNLO is
due to corrections to the coefficient $\tilde{c}_{ij}^{(2,0)}$ away
from the $m_t\to 0$ limit, and also to non-singular terms as $z\to 1$.
However, because the total cross section is dominated by values of $M$
where corrections in $m_t^2/M^2$ can be be significant, our analysis
does not allow us to distinguish the relative importance of the two.

\section{Conclusions}
\label{sec:conclusions}

We obtained an NNLO soft plus virtual approximation to the top-pair
invariant mass distribution at hadron colliders, valid up to
corrections of order $m_t^2/M^2$.  This is currently the most
complete approximation to the NNLO QCD corrections to a differential cross
section in top-quark pair production, and is most useful for the highly
boosted regime, where $m_t\ll M$.  
Nearly all the perturbative ingredients
needed for this calculation were available in the literature. However,
to extract the finite contribution from the NNLO virtual
corrections required us to implement a non-trivial IR subtraction
procedure and to calculate color-decomposed massless one-loop amplitudes to order $\epsilon^2$. We have not reprinted the rather
lengthy results here, instead giving them in {\sc Mathematica}  files which can
be downloaded from the source code of this paper available from 
the preprint server {\tt http://arXiv.org}.

We explored the phenomenological impact of our calculations in
Section~\ref{sec:pheno}.  We constructed an improved NNLO
approximation by adding our new results for non-logarithmic
(delta-function) corrections to the logarithmic (plus-distribution)
terms determined from NNLL soft gluon resummation for arbitrary $m_t$.
We observed that the new non-logarithmic corrections produce mild
enhancements of the differential cross section, which are
slightly more important at low values of invariant mass than at high
ones.  Implications for NNLO approximations to the total cross section
were also explored.  Finally, we compared the new NNLO approximation
with results from NNLL soft gluon resummation. At relatively low values of
invariant mass, where the cross section is large, resummation is only
a small effect and fixed-order perturbation theory is perfectly
sufficient.  On the other hand, for higher values of $M$ corrections
from NNLL resummation are quite large and should not be neglected, as
clearly illustrated in Figure~\ref{fig:M}. 

Several further things would be interesting to investigate. First, it
would be desirable to implement numerically the double resummation of
soft and small-mass logarithms, using the formalism developed in
\cite{Ferroglia:2012ku}.  The optimal prediction would combine these
with the NNLO calculations obtained here, or even better with the full
NNLO results once they become available. Second, given the importance
of soft-gluon resummation, it would be interesting to compare
numerical results from Mellin-space resummation
\cite{Kidonakis:1997gm} with the momentum-space results from
\cite{Ahrens:2010zv}.  Both types of resummation should also be
compared with results from parton shower codes.  Finally, electroweak
corrections start to become more significant at higher invariant mass
\cite{Kuhn:2006vh,Kuhn:2013zoa}, and could be combined with the QCD
corrections in a numerical code.

\section*{Acknowledgments}
We would like to thank Nigel Glover for providing us the results of
\cite{Anastasiou:2001sv} in electronic form.
The work of A.F.\ was supported in part by the PSC-CUNY Award No.\ 65214-00-43 and by the National Science Foundation Grant No.\ PHY-1068317. 

\newpage

\appendix

\section{Constructing the IR counterterms}
\label{sec:Zmatrices}

In this appendix we give explicit results for the IR counterterms in Eqs.~(\ref{eq:ct02}) and~(\ref{eq:ct11}). To do so, we describe the ingredients needed to evaluate matrix elements of the form $\braket{ \mathcal{M}_i | \bm{Z}| \mathcal{M}_j }$ ($i, j \in \{0,1\}$). We briefly review the color-space algebra underlying the braket notation, and then give perturbative results for the different elements needed in our analysis.  

So far, we have described the renormalization factor $\bm{Z}$ and the amplitudes $\ket{\mathcal{M}}$ in the color-basis independent notation of \cite{Catani:1996jh}. To calculate these for a specific process, one must first define a color basis. We are interested in the partonic processes $q^{a_1} \bar{q}^{a_2}, g^{a_1} g^{a_2} \to Q^{a_3} \bar{Q}^{a_4}$, where $a_i$ are the color labels of the partons involved in the scattering process. We denote the set of four color labels collectively as $\{a\}$ and define the $s$-channel singlet-octet basis \cite{Ferroglia:2012ku, Ferroglia:2012uy}:
\begin{gather}
  (c_1^{q \bar{q}})_{\{a\}} = \delta_{a_2 a_1} \delta_{a_3 a_4} \, , \quad (c_2^{q \bar{q}})_{\{a\}}  = t^c_{a_2 a_1} t^c_{a_3 a_4}  \, , \nonumber
  \\
  (c_1^{gg})_{\{a\}} =\delta^{a_2 a_1} \delta_{a_3 a_4} \, , \quad (c_2^{gg})_{\{a\}} = i f^{a_1 a_2 c} \, t^c_{a_3 a_4} \, , \quad (c_3^{gg})_{\{a\}} = d^{a_1 a_2 c} \, t^c_{a_3 a_4} \, .
  \label{eq:structures}
\end{gather}
The structures in Eq.~(\ref{eq:structures}) are the explicit forms of the basis vectors $\ket{c_I}$. Inner products in the color space can be calculated following
\begin{align}
  \braket{ c_I |  c_J } = \sum_{\{a\}} \left( c_I \right)^*_{\{a_1 a_2 a_3 a_4\}} (c_J)_{\{a_1 a_2 a_3 a_4\}} \, .
\end{align}
The basis vectors defined above are orthogonal but not orthonormal. For instance, the leading-order soft function has matrix elements $\tilde{s}_{IJ}^{(0)}=\braket{ c_I | c_J }/d_R$ and reads in the two channels: 
\begin{align}
  \label{eq:Stree}
  \tilde{\bm{s}}_{q\bar{q}}^{(0)} =
  \begin{pmatrix}
    N & 0
    \\
    0 & \frac{C_F}{2}
  \end{pmatrix}
  , \quad \tilde{\bm{s}}_{gg}^{(0)} = 
  \begin{pmatrix}
    N & 0 & 0
    \\
    0 & \frac{N}{2} & 0
    \\
    0 & 0 & \frac{N^2-4}{2N}
  \end{pmatrix}
  \,.
\end{align}
By employing the completeness relation in color space
\begin{align}
  \bm{1} = \sum_I \frac{1}{\braket{ c_I | c_I }}  \ket{c_I} \bra{c_I} \, , 
\end{align}
we can write the matrix elements appearing in the IR counterterms as
(suppressing sums over repeated indices)
\begin{align}
\langle {\mathcal M}_i |\bm{Z}| {\mathcal M}_j \rangle &=   \frac{1}{\langle c_I | c_I \rangle \langle c_J | c_J \rangle} \langle {\mathcal M}_i | c_I \rangle \langle c_I |\bm{Z}|
c_J \rangle \langle c_J |{\mathcal M}_j \rangle  \nonumber \\
& =\left( \frac{1}{\langle c_J | c_J \rangle} \langle c_J |{\mathcal M}_j \rangle  \langle {\mathcal M}_i | c_I \rangle  \right) \left( \frac{1}{\langle c_I | c_I \rangle}  \langle c_I |\bm{Z}|
c_J \rangle\right) \, .
\end{align}
Looking at the arrangement of the terms in the second line, it 
is natural to define  matrices $\bm{{\mathcal Z}}$ and 
$\bm{\mathcal{M}}^{(i,j)}$ with elements
\begin{align}
{\mathcal Z}_{IJ} \equiv \frac{1}{\langle c_I | c_I\rangle} \langle c_I | \bm{Z} | c_J\rangle \, , \qquad 
 {\mathcal M}^{(i,j)}_{IJ}\equiv\frac{1}{\langle c_I | c_I \rangle} \langle c_I |{\mathcal M}_i \rangle  \langle {\mathcal M}_j | c_J \rangle  \, ,
\end{align}
so that  
\begin{align} \label{eq:matr}
\langle {\mathcal M}_i |\bm{Z}| {\mathcal M}_j \rangle &= \Tr\left[ \bm{{\mathcal M}}^{(j,i)} \bm{{\mathcal Z}}\right] \, .
\end{align}

At this point, we have reduced the problem of evaluating the IR
counterterms to that of specifying the matrix elements of
$\bm{{\mathcal Z}}$ and $\bm{\mathcal{M}}^{(i,j)}$.  The matrix
elements of $\bm{{\mathcal Z}}$ can be obtained in a straightforward
way starting from the basis-independent $\bm{Z}$-matrix for a generic
$n$-parton process in massless QCD derived in \cite{Becher:2009cu,
  Becher:2009qa}.  We list the results up to NNLO at the end of this
appendix.  Obtaining the matrix elements of $\bm{\mathcal{M}}^{(i,j)}$ is more involved.  In particular, since the diagonal
elements of the NLO matrix $\bm{{\mathcal Z}}$ contain double poles in
the dimensional regulator $\epsilon$, obtaining the order $\epsilon^0$
contribution from terms such as $ \langle {\mathcal M}_0 | \bm{Z}_1|
{\mathcal M}_1 \rangle$ in Eq.~(\ref{eq:ct02}) or $\langle {\mathcal
  M}_0 | \bm{Z}^\dagger_1| {\mathcal M}_1 \rangle$ in
Eq.~(\ref{eq:ct11}) required us to calculate the NLO matrix $\bm{\mathcal{M}}^{(1,0)}$ to order $\epsilon^2$. We briefly describe
the calculational procedure below.  The leading-order matrix
$\bm{\mathcal{M}}^{(0,0)}$ involves only tree level amplitudes and
can be easily calculated exactly in $\epsilon$.  The analytic
expressions for the matrix elements of $\bm{\mathcal{M}}^{(1,0)}$ are
rather lengthy, therefore we decided to include them in computer files which
can be found in the arXiv submission of this paper.

The calculation of $\bm{{\mathcal M}}^{(1,0)}$ involves the
interference of tree-level and one-loop amplitudes.  It is very
similar the calculation of the one-loop corrections to massless $2 \to
2$ scattering carried out long ago in \cite{Ellis:1985er} up to order
$\epsilon^0$.  We must modify that calculation by projecting the
amplitudes onto the color bases in Eq.~(\ref{eq:structures}), thus
forming the matrix structure, and by evaluating the master integrals
to order $\epsilon^2$.  While the results are new, we have checked
them in the following way.  The quantity corresponding to the matrix
$\bm{{\mathcal M}}^{(1,0)}$ for the case of massive quarks was
evaluated up to order $\epsilon$ in \cite{Ferroglia:2009ii}, and was
used to predict the IR poles in the two-loop corrections to top pair
production. Up to order $\epsilon$, we could then cross check the
calculation of the massless matrix $\bm{{\mathcal M}}^{(1,0)}$ by taking the
$m_t \to 0$ limit of the massive calculation and verifying that the
factorization formula \cite{Mitov:2006xs} connecting the small-mass
limit of QCD amplitudes with massless ones is satisfied.  
A further check on the order $\epsilon^2$ pieces of the diagonal matrix elements
is provided by using this same factorization formula to reproduce the
two-loop virtual corrections in the small-mass limit obtained in
\cite{Czakon:2007ej, Czakon:2007wk} (ignoring the contributions of heavy-quark loops).

We end this appendix by providing explicit expressions for the
matrix elements ${\mathcal Z}_{IJ}$ in both production channels (we
also give these in electronic form in the arXiv submission).
With the normalization chosen, the $\bm{{\mathcal Z}}$ matrix at order in $\alpha_s^0$ coincides  with the identity matrix in both channels:
\begin{align}
 \bm{{\mathcal Z}} & =  \bm{1} + \frac{\alpha_s}{4 \pi} \bm{{\mathcal Z}}^{(1)} + \left(\frac{\alpha_s}{4 \pi}\right)^2 \bm{{\mathcal Z}}^{(2)} +  \cdots \, .
\end{align}
One should observe that for the matrix elements of $\bm{Z}^\dagger$ one finds
\begin{align}
{\mathcal Z}_{IJ}^\dagger   = 
\frac{\langle c_J | c_J\rangle}{\langle c_I | c_I\rangle} {\mathcal Z}_{JI}^* \,. 
\end{align}
Note that the inner products of the basis vectors needed to evaluate
such an expression can be read off from Eq.~(\ref{eq:Stree}).

In the quark annihilation channel the matrices for the $n$-th order correction have the form
\begin{align}
\bm{{\mathcal Z}}^{(n)}_{q \bar{q}}  &= \left(\begin{array}{cc}
{\mathcal Z}^{(n)}_{11} &  {\mathcal Z}^{(n)}_{12}\\ 
{\mathcal Z}^{(n)}_{21} & {\mathcal Z}^{(n)}_{22} 
\end{array}  \right) \, .
\end{align}
The NLO matrix elements are
\begin{align}
{\mathcal Z}^{(1)}_{11} & = -\frac{4 C_F}{\epsilon^2} - \frac{1}{\epsilon}\left( 6 C_F + 4 C_F  L_s  \right)\, , \nonumber \\
{\mathcal Z}^{(1)}_{12} & =  \frac{2 C_F}{\epsilon N} \left( L_u -L_t\right)\, , \nonumber \\
{\mathcal Z}^{(1)}_{21} & = \frac{2 N}{C_F} {\mathcal Z}^{(1)}_{12}\, , \nonumber \\
{\mathcal Z}^{(1)}_{22} & = 
-\frac{4 C_F}{\epsilon^2} + \frac{2}{\epsilon} \left[  
L_t \left(\frac{2}{N} -  N\right) + \frac{L_s}{N} - \frac{2 L_u}{N}-3 C_F\right] \, ,
\end{align}
where we introduced the following notation
\begin{align}
L_s = \ln\left(-\frac{\mu^2}{\hat{s}}\right)  = \ln\left(\frac{\mu^2}{\hat{s}}\right) + i \pi  \, , \quad L_t = \ln\left(-\frac{\mu^2}{t_1}\right)\, , \quad L_u = \ln\left(-\frac{\mu^2}{u_1}\right) \, .
\end{align}
The elements of the NNLO matrix are
\begin{align}
{\mathcal Z}^{(2)}_{11} & =  \frac{8 C_F^2}{\epsilon^4}  -\frac{2 C_F}{\epsilon^3} \left[ \frac{1}{N} \left( 6 + 4 L_s\right) -
N \left(\frac{23}{2} + 4 L_s \right) + N_l\right] 
 + \frac{C_F}{\epsilon^2} \left[ \frac{1}{N} \left( 4 (L_t -L_u)^2 \right. \right.\nonumber \\
&\left. \left. -9 - 12 L_s - 4 L_s^2 \right) + N \left( \frac{113}{9} +  \frac{58}{3} L_s  + 4 L_s^2 + \frac{\pi^2}{3}\right) - \frac{4 N_l}{3} \left( \frac{2}{3} + L_s\right)\right] \nonumber \\
& + \frac{C_F}{\epsilon} \left[ \frac{1}{N} \left( \frac{3}{4} - \pi^2 + 12 \zeta(3)\right) + N \left( -\frac{2003}{108} -  \frac{134}{9} L_s - \frac{5}{6} \pi^2 + 
   \frac{2}{3} \pi^2 L_s + 14 \zeta(3)\right) \right. \nonumber \\ 
   & \left. + N_l \left(\frac{65}{27} + \frac{20}{9} L_s + \frac{\pi^2}{3} \right) \right] \, ,
\nonumber \\
{\mathcal Z}^{(2)}_{12} & =  \frac{4 C_F}{\epsilon^3} \left(\frac{1}{N^2} -1\right) \left( L_u -L_t\right) +\frac{C_F}{\epsilon^2} 
\left[ \frac{2}{N^2} \left( -3 L_t - 2 L_s L_t - 2 L_t^2 + 3 L_u + 
   2 L_s L_u \right. \right.
\nonumber \\
&\left. \left.
   + 4 L_t L_u - 2 L_u^2\right) + \frac{29}{3} L_t + 2 L_s L_t + 2 L_t^2 - \frac{29}{3} L_u - 2 L_s L_u - 
2 L_t L_u +\frac{2}{3} \frac{N_l}{N} \left(L_u -L_t \right)\right] \nonumber \\
& + \frac{C_F}{\epsilon}\left[ \left( \frac{67}{9} -\frac{\pi^2}{3} \right) \left( L_u -L_t\right) - \frac{10}{9} \frac{N_l}{ N} \left(L_u -L_t \right)\right] \, , \nonumber \\
{\mathcal Z}^{(2)}_{21} & =\frac{2 N}{C_F} {\mathcal Z}^{(2)}_{12}\, , \nonumber \\
{\mathcal Z}^{(2)}_{22} & = \frac{8 C_F^2}{\epsilon^4} + \frac{C_F}{\epsilon^3} \left[ \frac{4}{N} \left(4 L_u- 4 L_t -3 - 2 L_s  \right)  + N \left( 23 + 8 L_t\right)   -2 N_l\right]  \nonumber \\ 
& + \frac{1}{\epsilon^2} \left[ \frac{1}{N^2} \left( \frac{9}{2} + 6 L_s + 2 L_s^2 + 12 L_t + 8 L_s L_t + 6 L_t^2 - 
   12 L_u - 8 L_s L_u - 12 L_t L_u + 6 L_u^2\right) \right. 
   \nonumber \\
& \left.
   -\frac{97}{9} - \frac{29}{3} L_s - \frac{76}{3} L_t - 4 L_s L_t - 6 L_t^2 + 
    \frac{58}{3} L_u  + 4 L_t L_u + 2 L_u^2  - \frac{\pi^2}{6} + N^2 \left( \frac{113}{18} + \frac{29}{3} L_t 
\right. \right. \nonumber \\
& \left. \left.
    + 2 L_t^2 + \frac{\pi^2}{6}\right) + \frac{N_l}{N} \left(\frac{4}{9} + \frac{2}{3} L_s + \frac{4}{3} L_t - \frac{4}{3} L_u \right) - N_l N \left( \frac{4}{9} + \frac{2}{3} L_t \right)\right] +\frac{1}{\epsilon} \left[ \frac{1}{N^2} \left( -\frac{3}{8} + \frac{\pi^2}{2} 
    \right. \right. \nonumber \\
    & \left. \left.
    - 6 \zeta(3)\right) + \frac{521}{54} + \frac{67}{9} L_s + \frac{134}{9} L_t - \frac{134}{9} L_u - 
     \frac{\pi^2}{12} - \frac{\pi^2}{3} L_s +\frac{2 \pi^2}{3} (L_u -L_t)  - \zeta(3) \right.
\nonumber \\
& 
\left.
+ N^2 \left( -\frac{2003}{216} - \frac{67}{9} L_t - \frac{5}{12} \pi^2 + 
  \frac{\pi^2}{3} L_t + 7 \zeta(3)\right) +\frac{N_l}{N} \left( -\frac{65}{54} - \frac{10}{9} L_s + \frac{20}{9} (L_u - L_t)
\right . \right. \nonumber \\
& 
\left. \left.
- \frac{\pi^2}{6}\right) + N_l N \left( \frac{65}{54} + \frac{10}{9} L_t +\frac{\pi^2}{6} \right)
     \right]\, . 
\end{align}

 In the gluon fusion channel, one deals with $3 \times 3$ matrices:
 \begin{align}
 \bm{{\mathcal Z}}^{(n)}_{gg}  &= \left(\begin{array}{ccc}
 {\mathcal Z}^{(n)}_{11} &  {\mathcal Z}^{(n)}_{12} &  {\mathcal Z}^{(n)}_{13}\\ 
 {\mathcal Z}^{(n)}_{21} & {\mathcal Z}^{(n)}_{22} &  {\mathcal Z}^{(n)}_{23} \\
  {\mathcal Z}^{(n)}_{31} & {\mathcal Z}^{(n)}_{32} &  {\mathcal Z}^{(n)}_{33}
 \end{array}  \right) \, .
 \end{align}
At NLO the matrix elements are 
\begin{align}
{\mathcal Z}^{(1)}_{11} & = \frac{1}{\epsilon^2} \left( \frac{1}{N} - 3 N\right) + \frac{1}{\epsilon } \left[ \frac{1}{N}\left( \frac{3}{2} + L_s\right) -N \left( \frac{31}{6} + 3 L_s \right) + \frac{2}{3} N_l \right] \, , \nonumber \\
{\mathcal Z}^{(1)}_{12} & = \frac{2}{\epsilon} \left( L_u -L_t\right)  \, , \nonumber \\
{\mathcal Z}^{(1)}_{13} & =  0\, , \nonumber \\
{\mathcal Z}^{(1)}_{21} & = 2 {\mathcal Z}^{(1)}_{12}\, , \nonumber \\
{\mathcal Z}^{(1)}_{22} & =   \frac{1}{\epsilon^2} \left( \frac{1}{N} - 3 N\right) + \frac{1}{\epsilon } \left[\frac{1}{N}\left( \frac{3}{2} + L_s\right) -N \left( \frac{31}{6} +  L_s  + L_t +L_u\right) + \frac{2}{3} N_l  \right]\, , \nonumber \\
{\mathcal Z}^{(1)}_{23} & =  \frac{1}{\epsilon} \left[ \frac{4}{N} \left( L_t -L_u\right) + N \left( L_u -L_t\right)\right] \, , \nonumber \\
{\mathcal Z}^{(1)}_{31} & = 0 \, , \nonumber \\
{\mathcal Z}^{(1)}_{32} & = \frac{N^2}{N^2-4}{\mathcal Z}^{(1)}_{23} \, , \nonumber \\
{\mathcal Z}^{(1)}_{33} & = \frac{1}{\epsilon^2} \left( \frac{1}{N} - 3 N\right) + \frac{1}{\epsilon }  \left[\frac{1}{N}\left( \frac{3}{2} + L_s\right) -N \left( \frac{31}{6} +  L_s  + L_t +L_u\right) + \frac{2}{3} N_l  \right]\, .
\end{align}
At NNLO one finds
\begin{align}
{\mathcal Z}^{(2)}_{11} & = \frac{1}{\epsilon^4}  \Biggl(\frac{9}{2} N^2 +\frac{1}{2 N^2}-3 \Biggr)  + \frac{1}{\epsilon^3}
\Biggl[\frac{1}{N^2}\left(\frac{3}{2} + L_s\right)
-\frac{149}{12} - 6 L_s +   
 N^2 \left(\frac{95}{4} + 9 L_s \right) \nonumber \\
 &+ \frac{7 N_l}{6 N} - \frac{7}{2} N_l N \Biggr] + \frac{1}{\epsilon^2} \Biggl[ 
 \left(\frac{9 }{2}L_s^2+21 L_s+\frac{\pi
    ^2}{4}+\frac{1241}{72}\right)
    N^2+\frac{1}{N^2}\left(\frac{1}{2}L_s^2+\frac{3}{2} L_s+\frac{9}{8}\right)-3
    L_s^2
\nonumber \\ 
&+N N_l \left(-3 L_s
    -\frac{50}{9}\right)+\frac{N_l}{N}\left(L_s
    +\frac{11 }{9}\right)-\frac{23 }{2}L_s+4
    L_t^2-8 L_t L_u+4
    L_u^2+\frac{4 }{9}N_l^2
\nonumber \\ 
&-\frac{\pi
    ^2}{12}-\frac{311}{36}
 \Biggr] + \frac{1}{\epsilon} \Biggl[ 
 N^2 \left(\frac{\pi ^2 }{2}L_s-\frac{67
   }{6} L_s+\frac{9 \zeta (3)}{2}+\frac{7 \pi
    ^2}{72}-\frac{2513}{144}\right)+N N_l
    \Biggr(\frac{5  }{3}L_s+\frac{\pi ^2
   }{36} 
\nonumber \\
&
   +\frac{125}{36}  \Biggl)-\frac{N_l}{N}\left(\frac{5 }{9}  L_s+\frac{\pi ^2}{12} +\frac{119}{108}\right)
   +\frac{67}{18}  L_s-\frac{\pi ^2}{6}  L_s-\frac{1}{N^2}\Biggl(3
    \zeta (3)+\frac{3}{16}-\frac{\pi
    ^2}{4}\Biggr)
\nonumber \\
& -\frac{\zeta (3)}{2}-\frac{\pi
    ^2}{24}+\frac{521}{108}
 \Biggr]
\, , \nonumber \\
{\mathcal Z}^{(2)}_{12} & = \frac{1}{\epsilon^3} \Biggl(\frac{2}{N} - 6 N \Biggr)(L_u- L_t) +  \frac{1}{\epsilon^2} \Biggl[N \left(4 L_s L_t-4 L_s
      L_u+L_t^2+14 L_t-L_u^2-14
      L_u\right) 
\nonumber \\ &
      -\frac{1}{N} \Biggl(2 L_s L_t-2
      L_s L_u+3 L_t-3
      L_u\Biggr)-2 L_t N_l+2
      L_u N_l\Biggr]+  \frac{1}{\epsilon} \Biggl[N \Biggl(\frac{\pi ^2}{3} L_t-\frac{67}{9} L_t+\frac{67 }{9}L_u
\nonumber \\ &
-\frac{\pi ^2}{3} L_u\Biggr)+\frac{10}{9} N_l L_t-\frac{10}{9} N_l L_u\Biggr]   \, , \nonumber \\
{\mathcal Z}^{(2)}_{13} & =  \frac{1}{\epsilon^2} \Biggl(N -\frac{4}{N}\Biggr)  (L_u-L_t)^2 \, , \nonumber \\
{\mathcal Z}^{(2)}_{21} & = 2 {\mathcal Z}^{(2)}_{12}   \, , \nonumber \\
{\mathcal Z}^{(2)}_{22} & = \frac{1}{\epsilon^4} \Biggl[\frac{9 }{2} N^2+\frac{1}{2 N^2}-3\Biggr] +\frac{1}{\epsilon^3} \Biggl[N^2 \left(3 L_s+3 L_t+3
   L_u+\frac{95}{4}\right)+\frac{1}{N^2}\left(L_s+\frac
   {3}{2}\right)-4
   L_s-L_t
\nonumber \\ &
   -L_u-\frac{7 N
   N_l}{2}+\frac{7 N_l}{6
   N}-\frac{149}{12}\Biggr] +  \frac{1}{\epsilon^2} \Biggl[N^2 \Biggl(\frac{L_s^2}{2}+L_s
      L_t+L_s L_u+7
      L_s+L_t^2+7 L_t+L_u^2+7
      L_u
\nonumber \\ &
      +\frac{\pi
      ^2}{4}+\frac{1241}{72}\Biggr)+ \frac{1}{N^2} \Biggl( \frac{1}{2}L_s^2+\frac{3}{2} L_s+\frac{9}{8}\Biggr)-L_s^2
      -N  N_l\left(L_s  + L_t + L_u+\frac{50}{9} \right)
\nonumber \\ &
      -L_s L_t-L_s
      L_u+\frac{N_l}{N} \Biggl(L_s+\frac{11}{9}\Biggr)-\frac{17}{2} L_s+2
      L_t^2-4 L_t L_u-\frac{3}{2} L_t+2 L_u^2-\frac{3}{2} L_u+\frac{4}{9} N_l^2
\nonumber \\ &
      -\frac{\pi
      ^2}{12}-\frac{311}{36}\Biggr]+  \frac{1}{\epsilon} \Biggl[N^2 \Biggl(\frac{\pi ^2}{6} L_s-\frac{67}{18} L_s-\frac{67 L_t}{18}+\frac{\pi ^2}{6} L_t-\frac{67}{18} L_u+\frac{\pi ^2}{6} L_u+\frac{9 \zeta (3)}{2}+\frac{7 \pi
         ^2}{72}
\nonumber \\ &
         -\frac{2513}{144}\Biggr)+N N_l
         \Biggl(\frac{5}{9} L_s+\frac{5}{9} L_t+\frac{5}{9}  L_u+\frac{\pi ^2}{36}+\frac{125}{36} \Biggr)-
         \frac{N_l}{N} \Biggl(\frac{5}{9} L_s+\frac{\pi ^2}{12}+\frac{119}{108}\Biggr)+\frac{67}{18}L_s
\nonumber \\ &
         -\frac{\pi ^2}{6} L_s - \frac{1}{N^2}\Biggl(3
         \zeta (3)+\frac{3}{16}-\frac{\pi^2}{4}\Biggr)-\frac{\zeta (3)}{2}-\frac{\pi
         ^2}{24}+\frac{521}{108}\Biggr]  \, , \nonumber \\
{\mathcal Z}^{(2)}_{23} & =  \frac{1}{\epsilon^3} \Biggl(3 N^2+\frac{4}{N^2}-13 \Biggr) ( L_t- L_u) +  \frac{1}{\epsilon^2} \Biggl[N^2 \left(L_s L_t-L_s
      L_u+L_t^2+7 L_t-L_u^2-7
      L_u\right)
\nonumber \\ &
      +\frac{1}{N^2} \Biggl(4 L_s L_t-4
      L_s L_u+6 L_t-6
      L_u\Biggr)-5 L_s L_t+5
      L_s L_u-4 L_t^2+\frac{N_l}{N} \left(4 L_t-4 L_u \right) 
\nonumber \\ &
      +N N_l (L_u-L_t)-\frac{59}{2}L_t+4 L_u^2+\frac{59}{2} L_u\Biggr]+  \frac{1}{\epsilon} \Biggl[N^2 \left(\frac{\pi ^2}{6} L_t-\frac{67}{18} L_t+\frac{67}{18} L_u-\frac{\pi ^2}{6} L_u\right)
\nonumber \\ &
      +N N_l\left(\frac{5}{9} L_t-\frac{5}{9} L_u\right)+\frac{N_l}{N} \Biggl(\frac{20}{9} L_u-\frac{20}{9} L_t\Biggr) +\frac{134}{9} L_t-\frac{2 \pi ^2}{3} L_t-\frac{134}{9} L_u +\frac{2 \pi^2}{3} L_u\Biggr]   \, , \nonumber \\
{\mathcal Z}^{(2)}_{31} & =  \frac{2N^2}{N^2-4} {\mathcal Z}^{(2)}_{13} \, , \nonumber \\
{\mathcal Z}^{(2)}_{32} & =  \frac{N^2}{N^2-4}{\mathcal Z}^{(2)}_{23} \, , \nonumber \\
{\mathcal Z}^{(2)}_{33} & =  \frac{1}{\epsilon^4} \Biggl[\frac{9 N^2}{2}+\frac{1}{2 N^2}-3\Biggr] +\frac{1}{\epsilon^3} \Biggl[N^2 \left(3 L_s+3 L_t+3
   L_u+\frac{95}{4}\right)+\frac{1}{N^2} \Biggl( L_s+\frac
   {3}{2} \Biggr) -4
   L_s-L_t
\nonumber \\ &
   -L_u-\frac{7}{2} N N_l+\frac{7 N_l}{6
   N}-\frac{149}{12}\Biggr] +  \frac{1}{\epsilon^2} \Biggl[N^2 \Biggl(\frac{L_s^2}{2}+L_s
      L_t
      +L_s L_u+7
      L_s+L_t^2+7 L_t+L_u^2+7
      L_u
\nonumber \\ &
      +\frac{\pi
      ^2}{4}+\frac{1241}{72}\Biggr)+\frac{1}{N^2} \Biggl(\frac{1}{2} L_s^2+\frac{3}{2} L_s+\frac{9}{8}\Biggr)-L_s^2
      -N  N_l\left(L_s+L_t+L_u+\frac{50}{9}\right)-L_s L_t
\nonumber \\ &
      -L_sL_u+\frac{N_l}{N} \Biggl( L_s+\frac{11}{9}\Biggr) -\frac{17}{2} L_s-2
      L_t^2+4 L_t L_u-\frac{3}{2} L_t-2 L_u^2-\frac{3}{2} L_u+\frac{4 }{9} N_l^2-\frac{\pi
      ^2}{12}
\nonumber \\ &
      -\frac{311}{36}\Biggr]+  \frac{1}{\epsilon} \Biggl[N^2 \left(\frac{\pi ^2 }{6} L_s-\frac{67}{18}L_s-\frac{67}{18} L_t+\frac{\pi ^2}{6} L_t -\frac{67}{18} L_u+\frac{\pi ^2}{6} L_u+\frac{9}{2} \zeta (3)+\frac{7 \pi^2}{72}-\frac{2513}{144}\right)
\nonumber \\ &
      +N N_l\left(\frac{5}{9} L_s+\frac{5}{9} L_t+\frac{5}{9} L_u+\frac{\pi ^2}{36}+\frac{125}{36}\right)-\frac{N_l}{N} 
\Biggl(\frac{5}{9} L_s +\frac{\pi ^2}{12}+\frac{119}{108}\Biggr)+\frac{67}{18} L_s -\frac{\pi ^2}{6} L_s
\nonumber \\ & -
\frac{1}{N^2} \left(3\zeta (3)+\frac{3}{16}-\frac{\pi^2}{4}\right)-\frac{\zeta (3)}{2}-\frac{\pi
         ^2}{24}+\frac{521}{108}\Biggr]   \, .
\end{align}


\begin{thebibliography}{99}

\bibitem{Aad:2012hg} 
  G.~Aad {\it et al.}  [ATLAS Collaboration],
  Eur.\ Phys.\ J.\ C {\bf 73}, 2261 (2013)
  [arXiv:1207.5644 [hep-ex]].

\bibitem{Baernreuther:2012ws} 
  P.~Baernreuther, M.~Czakon and A.~Mitov,
  Phys.\ Rev.\ Lett.\  {\bf 109}, 132001 (2012)
  [arXiv:1204.5201 [hep-ph]].

\bibitem{Czakon:2012pz} 
  M.~Czakon and A.~Mitov,
  JHEP {\bf 1301}, 080 (2013)
  [arXiv:1210.6832 [hep-ph]].

\bibitem{Czakon:2013goa} 
  M.~Czakon, P.~Fiedler and A.~Mitov,
  arXiv:1303.6254 [hep-ph].

\bibitem{Ahrens:2010zv} 
  V.~Ahrens, A.~Ferroglia, M.~Neubert, B.~D.~Pecjak and L.~L.~Yang,
  JHEP {\bf 1009}, 097 (2010)
  [arXiv:1003.5827 [hep-ph]].

\bibitem{Ferroglia:2012ku} 
  A.~Ferroglia, B.~D.~Pecjak and L.~L.~Yang,
  Phys.\ Rev.\ D {\bf 86}, 034010 (2012)
  [arXiv:1205.3662 [hep-ph]].

\bibitem{Ahrens:2009uz} 
  V.~Ahrens, A.~Ferroglia, M.~Neubert, B.~D.~Pecjak and L.~L.~Yang,
  Phys.\ Lett.\ B {\bf 687}, 331 (2010)
  [arXiv:0912.3375 [hep-ph]].

\bibitem{Ferroglia:2012uy} 
  A.~Ferroglia, B.~D.~Pecjak, L.~L.~Yang, B.~D.~Pecjak and L.~L.~Yang,
  JHEP {\bf 1210}, 180 (2012)
  [arXiv:1207.4798 [hep-ph]].

\bibitem{Melnikov:2004bm} 
  K.~Melnikov and A.~Mitov,
  Phys.\ Rev.\ D {\bf 70}, 034027 (2004)
  [hep-ph/0404143].

\bibitem{Chuvakin:2001ge} 
  A.~Chuvakin and J.~Smith,
  Comput.\ Phys.\ Commun.\  {\bf 143}, 257 (2002)
  [hep-ph/0103177].

\bibitem{Anastasiou:2000kg} 
  C.~Anastasiou, E.~W.~N.~Glover, C.~Oleari and M.~E.~Tejeda-Yeomans,
  Nucl.\ Phys.\ B {\bf 601}, 318 (2001)
  [hep-ph/0010212].

\bibitem{Anastasiou:2001sv} 
  C.~Anastasiou, E.~W.~N.~Glover, C.~Oleari and M.~E.~Tejeda-Yeomans,
  Nucl.\ Phys.\ B {\bf 605}, 486 (2001)
  [hep-ph/0101304].

\bibitem{Anastasiou:2000mv} 
  C.~Anastasiou, E.~W.~N.~Glover, C.~Oleari and M.~E.~Tejeda-Yeomans,
  Phys.\ Lett.\ B {\bf 506}, 59 (2001)
  [hep-ph/0012007].

\bibitem{Glover:2001af} 
  E.~W.~N.~Glover, C.~Oleari and M.~E.~Tejeda-Yeomans,
  Nucl.\ Phys.\ B {\bf 605}, 467 (2001)
  [hep-ph/0102201].

\bibitem{Glover:2001rd} 
  E.~W.~N.~Glover and M.~E.~Tejeda-Yeomans,
  JHEP {\bf 0105}, 010 (2001)
  [hep-ph/0104178].

\bibitem{Xu}
T.~Becher, M.~Neubert and G.~Xu,
JHEP {\bf 0708}, 030 (2008)
[arXiv:0710.0680[hep-ph]]


\bibitem{Higgs1}
V.~Ahrens, T.~Becher, M.~Neubert and L.~L.~Yang,
Phys. Rev. {\bf D 79}, 033013 (2009)
[arXiv:0808.3008[hep-ph]]


\bibitem{Higgs2}
V.~Ahrens, T.~Becher, M.~Neubert and L.~L.~Yang,
Eur. Phys. J. {\bf C 62}, 333 (2009)
[arXiv:0809.4283[hep-ph]]

\bibitem{Ahrens:2011mw} 
  V.~Ahrens, A.~Ferroglia, M.~Neubert, B.~D.~Pecjak and L.~-L.~Yang,
  JHEP {\bf 1109}, 070 (2011)
  [arXiv:1103.0550 [hep-ph]].

\bibitem{Catani:1996jh} 
  S.~Catani and M.~H.~Seymour,
  Phys.\ Lett.\ B {\bf 378}, 287 (1996)
  [hep-ph/9602277].

\bibitem{Becher:2009cu} 
  T.~Becher and M.~Neubert,
  Phys.\ Rev.\ Lett.\  {\bf 102}, 162001 (2009)
  [arXiv:0901.0722 [hep-ph]].

\bibitem{Becher:2009qa} 
  T.~Becher and M.~Neubert,
  JHEP {\bf 0906}, 081 (2009)
  [arXiv:0903.1126 [hep-ph]].

\bibitem{Catani:1998bh} 
  S.~Catani,
  Phys.\ Lett.\ B {\bf 427}, 161 (1998)
  [hep-ph/9802439].

\bibitem{Czakon:2007ej} 
  M.~Czakon, A.~Mitov and S.~Moch,
  Phys.\ Lett.\ B {\bf 651}, 147 (2007)
  [arXiv:0705.1975 [hep-ph]].

\bibitem{Czakon:2007wk} 
  M.~Czakon, A.~Mitov and S.~Moch,
  Nucl.\ Phys.\ B {\bf 798}, 210 (2008)
  [arXiv:0707.4139 [hep-ph]].

\bibitem{Campbell:2000bg} 
  J.~M.~Campbell and R.~K.~Ellis,
  Phys.\ Rev.\ D {\bf 62}, 114012 (2000)
  [hep-ph/0006304].


\bibitem{Martin:2009iq} 
  A.~D.~Martin, W.~J.~Stirling, R.~S.~Thorne and G.~Watt,
  Eur.\ Phys.\ J.\ C {\bf 63}, 189 (2009)
  [arXiv:0901.0002 [hep-ph]].



\bibitem{Ahrens:2011px} 
  V.~Ahrens, A.~Ferroglia, M.~Neubert, B.~D.~Pecjak and L.~L.~Yang,
  Phys.\ Lett.\ B {\bf 703}, 135 (2011)
  [arXiv:1105.5824 [hep-ph]].

\bibitem{Kidonakis:1997gm}
  N.~Kidonakis and G.~F.~Sterman,
  Nucl.\ Phys.\ B {\bf 505} 321 (1997)
  [hep-ph/9705234].

\bibitem{Kuhn:2006vh} 
  J.~H.~K\"uhn, A.~Scharf and P.~Uwer,
  Eur.\ Phys.\ J.\ C {\bf 51}, 37 (2007)
  [hep-ph/0610335].

\bibitem{Kuhn:2013zoa} 
  J.~H.~K\"uhn, A.~Scharf and P.~Uwer,
  arXiv:1305.5773 [hep-ph].


\bibitem{Ellis:1985er} 
  R.~K.~Ellis and J.~C.~Sexton,
  Nucl.\ Phys.\ B {\bf 269}, 445 (1986).

\bibitem{Ferroglia:2009ii} 
  A.~Ferroglia, M.~Neubert, B.~D.~Pecjak and L.~L.~Yang,
  JHEP {\bf 0911}, 062 (2009)
  [arXiv:0908.3676 [hep-ph]].

\bibitem{Mitov:2006xs} 
  A.~Mitov and S.~Moch,
  JHEP {\bf 0705}, 001 (2007)
  [hep-ph/0612149].



\end{thebibliography}
\end{document}